\documentclass[sigconf]{acmart}
\AtBeginDocument{%
  }
\setcopyright{acmlicensed}
\copyrightyear{2026}
\acmYear{2026}
\acmDOI{XXXXXXX.XXXXXXX}
\acmConference[Submitted to ACM Middleware 2026]{Pre-Print}{XXXXX}{XXXXX}
\acmISBN{Pre-Print}

\usepackage{listings}
\usepackage{xcolor}
\usepackage{seqsplit}
\usepackage[scale=0.8]{FiraMono}
\usepackage{listings}
\usepackage{caption}
\usepackage{subcaption}
\usepackage{pgfplots}
\pgfplotsset{compat=1.18}
\usepackage{pgfplotstable}
\usepackage{subcaption}

\usepackage{newfloat}
\DeclareFloatingEnvironment[fileext=lol, name=Code Example]{listing}
\DeclareCaptionSubType{listing}

\lstdefinestyle{nsb}{
    basicstyle=\ttfamily\small,
    keywordstyle=\bfseries,
    commentstyle=\itshape,
    showstringspaces=false,
    breaklines=true,
    columns=fullflexible,
    keepspaces=true,
    xleftmargin=8pt,
    aboveskip=6pt,
    belowskip=6pt,
    captionpos=b,
}

\begin{document}

\title{NSB2: An Open-Source Modular Pipeline for Application-Network Co-Simulation}

\author{Harikrishna Kuttivelil}
\email{hkuttive@ucsc.edu}
\orcid{0000-0002-3186-6590}
\affiliation{%
  \institution{University of California, Santa Cruz}
  \city{Santa Cruz}
  \state{California}
  \country{USA}
}
\author{Lakshmi Krishnaswamy}
\email{lakrishn@ucsc.edu}
\orcid{0009-0004-1042-2239}
\affiliation{%
  \institution{University of California, Santa Cruz}
  \city{Santa Cruz}
  \state{California}
  \country{USA}
}
\author{Katia Obraczka}
\email{obraczka@ucsc.edu}
\orcid{0000-0001-9720-1690}
\affiliation{%
  \institution{University of California, Santa Cruz}
  \city{Santa Cruz}
  \state{California}
  \country{USA}
}
\author{Srikar Rani}
\affiliation{%
  \institution{University of California, Santa Cruz}
  \city{Santa Cruz}
  \state{California}
  \country{USA}
}
\author{Psi Padhya}
\affiliation{%
  \institution{University of California, Santa Cruz}
  \city{Santa Cruz}
  \state{California}
  \country{USA}
}
\author{Alex Woelkers}
\affiliation{%
  \institution{University of California, Santa Cruz}
  \city{Santa Cruz}
  \state{California}
  \country{USA}
}
\author{Vasavi Doddamani}
\affiliation{%
  \institution{University of California, Santa Cruz}
  \city{Santa Cruz}
  \state{California}
  \country{USA}
}
\author{Nayan Sanjay Bhatia}
\orcid{0009-0001-3314-7928}
\affiliation{%
  \institution{University of California, Santa Cruz}
  \city{Santa Cruz}
  \state{California}
  \country{USA}
}









\renewcommand{\shortauthors}{Kuttivelil, Krishnaswamy et al.}

\begin{abstract}
    As connected applications—from intelligence-enabled assistants and the smart homes to autonomous vehicles and large-scale Internet-of-Things deployments—become increasingly pervasive, their performance, safety, and reliability depend more than ever on the underlying network.
    Designing and validating such systems therefore requires development environments that holistically account for both application and network behavior.
    While network simulators and emulators provide high-fidelity models of network behavior, connecting full-fledged applications to them remains difficult due to platform constraints, simulator-specific interfaces, and narrow single-purpose integrations.
    To address this gap, we present NSB2 (Network Simulation Bridge 2), an open-source network co-simulation middleware that relays payloads between application processes through a network simulator.
    NSB2 is composed of a central daemon and thin, embedded application- and simulator-side libraries for flexible integration, disentangled communication channels for unencumbered inter-process communications, and in-memory key-value storage for efficient payload caching and retrieval.
    NSB2 is designed to be lightweight, accessible, and extensible: it exposes streamlined interfaces supported in Python and C++, modular internal abstractions that let users swap or extend network simulation/emulation backends, flexible and configurable deployment modes, and integration helpers for the widely-used \textit{ns-3} and \textit{OMNeT++} network simulators.
    Our evaluation across multiple co-simulation scenarios shows that NSB2 introduces negligible memory and latency overhead, enabling researchers and developers to achieve richer, more realistic modeling of connected systems with minimum overhead.
\end{abstract}



\maketitle

\newcommand{\citehere}{\color{orange}[CITE]\color{black}}
\newcommand{\notehere}[1]{\color{orange}[NOTE:\ #1]\color{black}}
\newcommand{\code}[1]{%
  {\fontseries{mb}\ttfamily%
    \hyphenchar\font=`\-%
    \StrSubstitute{#1}{.}{.\allowbreak}%
  }%
}
\newcommand{\comms}[1]{\textbf{\textsc{\MakeLowercase{#1}}}}
\newcommand{\config}[1]{\textit{\textsc{#1}}}

\section{Introduction}

Connected applications -- from intelligence-enabled assistants and the smart home to autonomous vehicles and large-scale Internet-of-Things deployments -- are no longer simple, point-to-point exchanges over a transparent communication layer.
These applications' performance, safety, and reliability critically depend on the behavior of the underlying network, including its latency, throughput, loss patterns, mobility, and contention dynamics.
As a result, the network is increasingly becoming a significant design factor, and the the development of such connected systems must holistically account for both the application logic and the network it runs on.
However, designing and validating connected applications and systems under realistic network conditions is fundamentally hard.
Physical testbeds offer some of the most realistic solutions through material hardware, but are expensive, brittle, and difficult to adapt to a diverse range of topologies and conditions.
Network simulators and emulators -- e.g., \textit{ns-3} \cite{ns3} and \textit{OMNeT++} \cite{omnet} -- provide high-fidelity, controllable, and reproducible models of network behavior.
However, connecting full-fledged applications to these simulators typically involves re-implementing limited proxies of the application's logic within the simulator runtime, utilizing privileged or platform-specific tunneling tools like TAP/TUN that are not widely available, or building bespoke solutions that are tightly coupled to a narrow set of usage scenarios.
Researchers and developers of connected systems thus have to choose between whether to pursue application realism or network realism, while modeling modern connected systems typically requires both.
To address this challenge, we turn towards network \textit{co-simulation} -- running application logic natively while a separate simulator models communication via simulated networks.

In this paper, we present \textbf{NSB2} -- the Network Simulation Bridge 2 -- a network co-simulation middleware that provides a daemon service and thin, embedded application- and simulator-side libraries to enable co-simulation between connected applications and network simulators, enabling users to prototype their applications over simulated networks.
NSB2 is designed to be \textit{lightweight}, \textit{accessible}, and \textit{extensible} -- it imposes minimal overhead on the applications and systems it serves, integrates seemlessly with mainstream simulators and operating systems with very few external dependencies, and provides well-defined interfaces and abstractions that users can adapt to fit their needs.
NSB2 builds on an earlier proof-of-concept of the Network Simulation Bridge~\cite{nsb} (discussed in \S\ref{sec:old-version-comp}); this paper presents a major redesign and re-implementation of the system, from endpoint to endpoint, to make it general-purpose, scalable, easy-to-use, and available for broad use as an open-source tool.
Specifically, our contributions are as follows:
\begin{itemize}
    \item \textbf{A low-overhead, low-latency co-simulation tool.} \newline
    {\small NSB2 provides a bridge consisting of a central daemon and lightweight endpoint clients, persistent communication channels with sub-channels for control- and data-plane disentanglement, versatile messaging protocols to facilitate system operation across the bridge, and an in-memory key-value store to offload payload content such that it does not hamper relay paths.}
    \item \textbf{Streamlined, symmetric, language-agnostic interfaces.} \newline
    {\small NSB2's both endpoint libraries expose small APIs -- socket-like \code{send} and \code{receive} on the application side and HTTP-like \code{fetch} and \code{post} on the simulator side -- implemented with full feature parity across Python and C++ and designed to minimize the scope of integration into application and simulator code.}
    \item \textbf{Modular internal abstractions for extensibility.} \newline
    {\small Modular backends (e.g., message transport and payload caching) are exposed as abstract base classes with function contracts, such that users can swap default components or implement new ones, fully within the scope of the module and without touching the rest of the system.}
    \item \textbf{Flexible and configurable deployment modes.} \newline
    {\small Among other features, NSB2 supports both polling and event-driven system operation, both system-wide and per-node simulator integrations, enabling it to fit a range of usage scenarios.}
    \item \textbf{First-class helpers for widely-used network simulators.} \newline
    {\small NSB2 provides integration templates for both the \textit{ns-3} and \textit{OMNeT++} network simulators, making the bridge directly usable with the two network simulation platforms that dominate open-source network research and education.}
    \item \textbf{Empirical NSB2 performance evaluation.} \newline
    {\small We test NSB2 across multiple co-simulation scenarios and demonstrate its low latency and minimal resource utilization, validating its design goal of being lightweight.}
\end{itemize}
NSB2 is open-source and available to the public\footnote{\textit{Public repository link omitted due to double-blind submission requirement.}}, including technical documentation and guides.

\paragraph*{Roadmap}
The remainder of the paper is organized as follows.
\S\ref{sec:background} situates NSB2 within the broad landscape of network simulation and co-simulation tools and differentiates it from the older version of NSB2.
\S\ref{sec:framework} details the NSB2 framework and architecture -- its components, communication and data infrastructure, configuration, and operational lifecycle.
\S\ref{sec:evaluation} presents our performance evaluation and validation of NSB2.
\S\ref{sec:conclusion} concludes the paper with a discussion of the available public repository, ongoing work, and future directions.
\section{Background \& Related Work} \label{sec:background}
Testing, validating and evaluating networked applications have traditionally relied on network simulators and emulators, hardware testbeds or a combination thereof.
In this section, we start by outlining notable open-source network simulation platforms followed by highlighting how NSB2 completely redesigns and re-implements the preliminary NSB's prototype architecture towards  more robust, scalable, extensible and efficient open-source system.


\subsection{Network Simulation \& Co-Simulation}

\textit{ns-3} and \textit{OMNeT++} are two of the most popular open-source discrete-event network simulators and have been widely used in the research and practitioner communities for simulating complex network stacks.
Modeling application traffic in network simulators often involves the use of traffic generators, which can greatly limit the fidelity of application models.
\textit{ns-3}~\cite{ns3} does provide support for injecting application processes into the network backend via the TAP interface.
However, the main drawback of this approach is its limited availability as it currently remains available only on Linux kernels and does not introduce any cross-platform alternatives and therefore is not platform-agnostic and there is no mechanism for the direct integration of unmodified applications with the simulator. 
\textit{OMNeT++}~\cite{omnet} offers a similar interface through its INET framework; however, the implementation also relies on the TAP interface and faces the same constraints.
\textit{Mininet}~\cite{mininet} is a network emulator that is widely used in the network research, practitioner. and education communities.
It exposes realistic interfaces to user-configurable network topolologies such that users can plug in their real applications.
However, \textit{Mininet} and its variants~\cite{mininet-wifi} provide limited network configuration options, making it significantly less useful when working with complex network models.


\paragraph*{Network Co-Simulation}\
Beyond network simulators, network co-simulation tools, such as \textit{nsclick}~\cite{nsclick} and \textit{Dockemu}~\cite{dockemu}, were developed to bridge applications to a network simulator backend, but themselves faced significant platform constraints.
With the \textit{nsclick} tool, the bridging was strictly applicable to the older \textit{ns-2} network simulator and purely based on Linux kernel-level interfaces.
\textit{Dockemu}~\cite{dockemu} and \textit{Dockemu 2.0}~\cite{dockemu2}, on the other hand, provide a robust all-in-one network co-simulation option, but heavily depend on Docker to orchestrate container-based nodes, limiting its usability in non-containerized application environments.
Unlike these and other co-simulation options that depend on tightly integrated technology stacks or closed environments, NSB2 is a developer tool that provides lightweight, platform-agnostic, simulator-agnostic middleware that can scale and support diverse connected application scenarios.

\subsection{Beyond the Old Bridge} \label{sec:old-version-comp}

While NSB2 builds on some foundational ideas of its previous version~\cite{nsb} -- primarily the server-client bridge architecture -- NSB2 is a complete redesign and re-implementation, i.e., no code from the previous version is present or used in NSB2.
More specifically, NSB2 provides several unique contributions and advantages as a result of its re-engineering:
\begin{itemize}
    \item \textit{Robust configuration options to support diverse usage scenarios.}
    While the previous version provided a fixed polling-based system and necessitated separate simulator clients per node, NSB2 offers support for various configuration operations for system mode, simulator mode, and payload caching (see \S\ref{sec:configuration}).
    \item \textit{Multi-channel transport for unobstructed client-daemon communication.}
    While the previous version used single TCP socket connections per client--daemon channel to facilitate all control- and data-plane messages, NSB2 uses multiple, purposeful sub-channels to disentangle transmission, reception, and control operations (see \S\ref{sec:framework-comms}).
    \item \textit{Payload caching for efficient relays.}
 Although the previous version relayed full payloads over the bridge and consequently incurred performance degradation with large payloads that needed to be chunked, NSB2 uses an in-memory key-value store to cache large payloads and track them with a payload key (see \S\ref{sec:framework-database}).
    \item \textit{Extensibility through versatile stack components and modular abstractions.}
    While the previous version used a fixed, tightly integrated technology stack, NSB2 leverages widely available and versatile stack components, with many of those stack components being implemented as plug-and-play modules (see \S\ref{sec:modular-interfaces}).
\end{itemize}
Other advantageous differentiating features include dynamic initialization handshakes to prevent inconsistent configuration, full Python and C++ feature parity and the endpoints through a single Protobuf schema, mature package installation and configuration, and removal of unnecessary communication exchanges.
Beyond the unique technical advantages NSB2 provides over the prior version, these changes also make NSB2 more conducive to adoption by the open-source community.

\section{NSB2 Framework \& Architecture} \label{sec:framework}
\begin{figure*}
    \centering
    \includegraphics[width=0.9\linewidth]{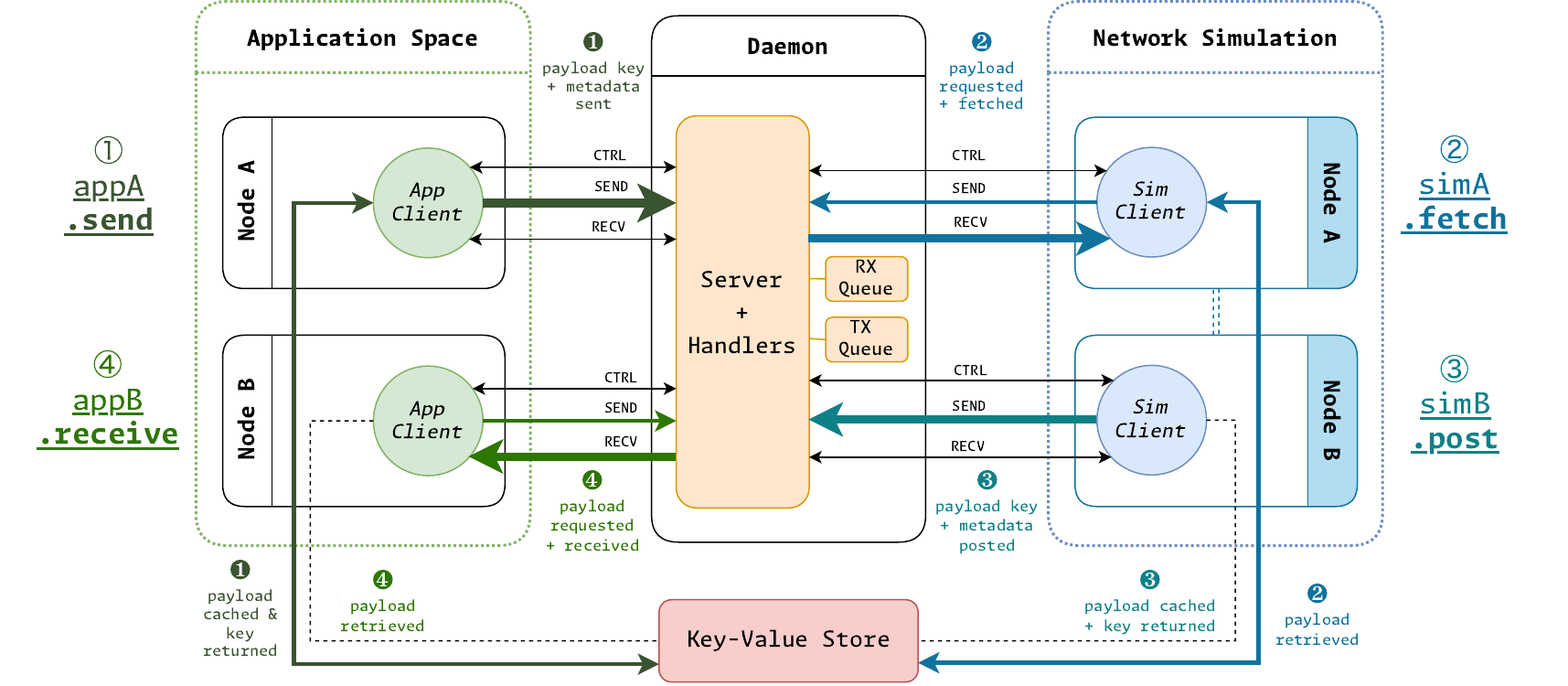}
    \caption{Lifecycle of a payload sent between application nodes.}
    \label{fig:lifecycle}
\end{figure*}

At its core, NSB2 is a high-bandwidth, low-latency message relaying framework.
In this section, we discuss the design of NSB2's pipeline, the roles of its components, and the implementation of its infrastructure.
An overview of the NSB system is presented in Figure~\ref{fig:lifecycle}.
We will mention the system configuration throughout this section; we cover it in detail in \S\ref{sec:configuration}.

\paragraph*{Design Goals}
NSB2 was designed to function as middleware for the development of distributed applications.
It serves as a simple, low-overhead co-simulation tool that bridges networked applications with network simulators, facilitating testing, evaluation and validation of these applications in realistic networked environments.
To this end, NSB2 has been designed to be: \textbf{lightweight}, \textbf{accessible}, \textbf{modular} and \textbf{extensible}.
We define \textit{lightweight} as resource-efficient, functionally minimal, and fast, such that it does not impose unnecessary burden on users, their simulations, or their systems.
This manifests in NSB2 as minimizing latency, memory utilization, and use of  external tools.
To strengthen \textit{accessibility}, NSB2 utilizes tools and dependencies that are available on most widely used platforms, including Linux (including Windows through WSL) and MacOS.
It is designed to download, install, and integrate on users' platforms with ease and transparency.
Complementary to its functional minimalism, NSB2 was designed to be \textit{extensible} such that users and contributors can innovate atop our framework through modular interfaces (see \S\ref{sec:modular-interfaces}).
NSB2 is already fully open-source, providing programming interfaces in Python and C++ to integrate into users' systems.

\paragraph*{Terminology}
In this paper, we use the term \textit{payload} to describe data being sent between application nodes and through the simulated network.
We use the term \textit{NSB message} -- or sometimes more simply, \textit{message} -- to describe system-level messages that facilitate inter-component communication and system operations.
Some of these \textit{messages} themselves carry \textit{payloads} as part its operation, as discussed in \S\ref{sec:framework-comms}.

\subsection{Framework Components} \label{sec:framework-components}
The NSB2 pipeline consists of endpoints -- the \textit{application client} and the \textit{simulator client} -- and a central coordinator -- the \textit{daemon}.
The daemon is the only stateful, long-running component while the endpoint clients are thin libraries embedded into the user's application or simulator.

\subsubsection{Daemon}
The daemon is a single, long-running process that centrally maintains the NSB2 pipeline.
It acts as a message broker -- instead of interacting directly with payloads, the application(s), or the simulator, it operates between the application and simulator clients to construct and coordinate the distributed system, to disambiguate communication and mediate system functions, and to direct the flow of payload data and metadata.
It owns a client registry with an entry per client containing the client's role and connection information and maintains transmission (TX) and reception (RX) buffers to keep track of payloads moving across the bridge.
Throughout the life of the system, the daemon also maintains the control plane of the system, enforcing the user configuration and allowing for system level functions like initialization, liveness checking, and shutdown.
During system initialization, the daemon is responsible for taking in the user's configuration specifications and initializing the system accordingly, then helping endpoint clients connect to the system (see \S\ref{sec:framework-system-init}).
During system runtime, the daemon acts a server responding to client requests to send and receive payloads, forwarding payloads as necessary when the system is in \config{PUSH} mode.
Notably, the daemon does not hold the payload content -- it handles \textit{payload entries} that contain payload keys and metadata, storing them in the TX/RX buffers.
It never accesses the inter-component messages beyond the control header and metadata for operational and routing purposes.

\subsubsection{Clients}
The application and simulator clients are thin wrapper libraries for system functions embedded within the user's application or simulator.
Clients do not communicate with each other directly and only communicating to the daemon through their multi-channel connection.
Both clients share similar functional mechanisms, such as initialization handshakes with the daemon, caching and retrieval of payloads.
Like the daemon, they use \textit{payload entries} to manage payloads, but can access and expose the full payload instead of just the key and metadata, so that the application and simulator logic can interface with the payload directly.
The minimal per-role logic and respective APIs are built around this shared transport and caching infrastructure.

\paragraph*{Application Client}
The application client -- \code{NsbAppClient} -- is the application-facing endpoint.
It presents a socket-like surface such that porting existing application to NSB2 only requires swapping I/O calls and can be unaware of the bridge, simulator, or wire protocol.
It has three API functions:
\begin{itemize}
    \item \code{send(destination, payload)}
    \newline {\small Transmission method to send the payload to the destination over the simulated network.}
    \item \code{receive()}
    \newline {\small Non-blocking polling method to check for and retrieve payloads received over the simulated network.}
    \item \code{listen(...)}
    \newline {\small Asynchronous callback-style implementation of \code{receive} to be used when the system is configured to be in \config{PUSH} mode.}
\end{itemize}
The application client can only see payloads and their metadata; they have no visibility of the network simulated in the network simulator and can only interact with it through these functions.

\paragraph*{Simulator Client}
The simulator client -- \code{NsbSimClient} -- is the simulator-facing endpoint providing hooks for the simulated network to be integrated into the overall payload pipeline.
It also has three complementary API functions:
\begin{itemize}
    \item \code{fetch()}
    \newline {\small\textit{(implemented within the transmission logic of the simulator)}}
    \newline {\small Non-blocking polling method to check for and retrieve payloads to be transmitted over the simulated network.}
    \item \code{listen(...)}
    \newline {\small Asynchronous callback-style implementation of \code{fetch} to be used when the system is configured to be in \config{PUSH} mode.}
    \item \code{post(payload, source, destination)}
    \newline {\small\textit{(implemented within the reception logic of the simulator})}
    \newline {\small Announcement method to notify delivery of the \textit{payload} from \textit{source} to \textit{destination}}
\end{itemize}
The number of instances of \code{NsbSimClient} is controlled by the specified \textit{simulator mode} in the configuration (see \S\ref{sec:configuration}).
Note that the functions listed above are largely symmetric within system operation, with endpoint-facing functions and APIs being specific to their role.
\code{NsbAppClient.send} and \code{NsbSimClient.post} both cache payloads and transmit payload metadata to the daemon over the \comms{SEND} channel.
\code{NsbSimClient.fetch} and \code{NsbSimClient.receive} both poll the daemon over the \comms{SEND} channel for outstanding payloads.
How these components and their functions come together to form the NSB2 pipeline is detailed in \S\ref{sec:framework-lifecycle}.

\paragraph*{Helper Modules for \textit{ns-3} \& \textit{OMNeT++}}
NSB2 provides specialized templates and packages tailored to the two most popular open-source network simulators -- \textit{ns-3} and \textit{OMNeT++} -- to ease integration.
The \textit{ns-3} template leverages the \config{system-wide} simulator configuration to allow users to implement traffic generation by using NSB2 to source traffic from payloads from the application space pending transmission.
For \textit{OMNeT++}, templates are provided for both the \config{per-node} configuration -- implementing NSB2 connectors as module-level applications -- and the \config{system-wide} configuration -- using its schedulers generate traffic using NSB2, similar to \textit{ns-3}.

\subsection{Communications \& Data Infrastructure} \label{sec:framework-infra}

The framework components are supported by infrastructure designed to keep NSB2 fast, lightweight, and extensible, including multi-channel communication architecture, an in-memory key-value database, and abstracted module interfaces.

\subsubsection{Multi-Channel Communication Architecture} \label{sec:framework-comms}


Each client in NSB2 -- whether it is \code{NsbAppClient} or \code{NsbSimClient} -- is initialized with a unique and persistent channel to the daemon.
This endpoint-to-daemon channel is composed of three independent connections, or \textit{sub-channels}, to serve different system operations (see Table~\ref{tab:protobuf-manifest-op}):
\begin{itemize}
    \item \comms{CTRL}: a bidirectional sub-channel to serve control messages
    \item \comms{SEND}: a unidirectional sub-channel to facilitate transmission of outgoing messages containing payloads
    \item \comms{RECV}: a bidirectional sub-channel to facilitate reception or listening of incoming messages containing payloads
\end{itemize}
The separation of each point-to-point channel into these sub-channels follows the principle of \textit{ontvlechten}, or disentanglement.
This creates dedicated channels for the control plane and incoming and outgoing data planes to avoid the head-of-line blocking and volatile communication performance under heavy load that would occur if a single connection was used to accommodate all traffic.
This disentanglement enables clients to handle parallel operations, e.g., listening for incoming messages on the \comms{RECV} channel while still being able to transmit outgoing messages on the \comms{SEND} channel.
The full separation of the control and data planes also enables the daemon to execute system-level functions independently of ongoing message relaying taking place on its clients' connections' \comms{SEND} and \comms{RECV} channels.

\subsubsection{Payload Caching}\label{sec:framework-database}

If every payload was passed wholly through the bridge along with the enveloping \textit{NSB Message} as done in the prior version of NSB~\cite{nsb}, then payloads that are greater than the host environment's buffer size will result in messages being chunked at each leg of the NSB pipeline incurring additional overhead in computation, communication bandwidth, and consequently latency, which is antithetical to the stated design goals of being lightweight and fast.
To circumvent this problem, NSB2 can be configured to leverage fast, in-memory key-value storage to minimize the impact of payload size on system performance.
When a payload enters the NSB2 pipeline -- i.e., when \code{NSBAppClient.send} is called -- it is stored in the database and the key to access the payload is returned.
This key ($<$32 B) is then used in lieu of the payload as it is transmitted through the bridge.
The endpoints that check out the payload -- e.g., \code{NSBAppClient.receive} and \code{NSBSimClient.fetch} -- use this key to retrieve the full payload from the database.
In this manner, we abstract the actual contents, and therefore its size, from the communication bridge while providing access to the payload and its contents where necessary.

\subsubsection{Message Protocols \& Operations} \label{sec:framework-messaging}
All communication between the framework components over their communication channels is carried out by a single, flexible \textit{protocol buffer}~\cite{protobuf} -- the \textit{NSB Message}.
The \textit{NSB Message} encodes all operational semantics into a unified structure with three top-level fields -- the \code{Manifest} and the \code{Metadata} -- $<$32 B, together -- and the variable-length message body.

\paragraph*{Manifest}
The \code{Manifest} serves as the control header for every \textit{NSB Message} and contains three sub-fields.
The \textit{Operation} sub-field (\code{op}) specifies the operation being performed; operations are described in Table~\ref{tab:protobuf-manifest-op}.
The \textit{Originator} sub-field (\code{og}) specifies the originator of the message -- \code{DAEMON}, \code{APP\_CLIENT}, \code{SIM\_CLIENT} -- which the daemon uses to disambiguate behavior on operations.
The \textit{Operational Code} sub-field (\code{code}) specifies the semantic status of the message; the meaning of the codes depend on the \textit{Originator} specified as described in Table~\ref{tab:protobuf-manifest-code}.
The \textit{Manifest} as a whole determines how the rest of the \textit{NSB Message} is parsed and processed when it is received by any of the framework components.

\begin{table*}[ht]
\centering
\caption{\textit{NSB Message} Protocol Operations -- \texttt{Manifest.op}}
\label{tab:protobuf-manifest-op}
\begin{tabular}{lccc p{0.5\linewidth}}
\hline
\textbf{Operation} & \textbf{Originator} & \textbf{Channel} & \textbf{Direction} & \textbf{Description} \\
\hline
\code{PING} & \code{*\_CLIENT} & \comms{CTRL} & Client $\leftrightarrow$ Daemon & \small{Liveness check. Daemon echoes back with \code{SUCCESS} or \code{FAILURE}.} \\
\code{INIT} & \code{*\_CLIENT} & \comms{CTRL} & Client $\leftrightarrow$ Daemon & \small{Registration handshake. Client sends \code{IntroDetails} (identifier and channel ports); daemon responds with \code{ConfigParams}.} \\
\code{SEND} & \code{APP\_CLIENT}  & \comms{SEND} & Client $\to$ Daemon & \small{Application client submits a payload to the daemon for queuing. No acknowledgement is returned.} \\
\code{FETCH} & \code{SIM\_CLIENT} & \comms{RECV} & Client $\leftrightarrow$ Daemon & \small{Simulator client polls for a queued outbound payload. Daemon responds with \code{MESSAGE} or \code{NO\_MESSAGE}. Used in PULL mode.} \\
\code{POST} & \code{SIM\_CLIENT} & \comms{SEND} & Client $\to$ Daemon & \small{Simulator client delivers a payload that has completed traversal of the simulated network. No acknowledgement is returned.} \\
\code{RECEIVE} & \code{APP\_CLIENT} & \comms{RECV} & Client $\leftrightarrow$ Daemon & \small{Application client polls for a delivered payload. Daemon responds with \code{MESSAGE} or \code{NO\_MESSAGE}. Used in PULL mode.} \\
\code{FORWARD} & \code{DAEMON} & \comms{RECV} & Daemon $\to$ Client & \small{Daemon proactively pushes a payload to a client without a prior request. Used exclusively in PUSH mode; the body and metadata are identical to the originating \code{SEND} or \code{POST} message.} \\
\code{EXIT} & \code{*\_CLIENT} & \comms{CTRL} & Client $\to$ Daemon & \small{Signals graceful shutdown. The daemon stops on receipt; no response is sent.} \\
\hline
\end{tabular}
\end{table*}

\begin{table}[!ht]
\centering
\caption{\textit{NSB Message} Status Codes -- \texttt{Manifest.code}}
\label{tab:protobuf-manifest-code}
\begin{tabular}{l p{0.7\linewidth}}
\hline
\textbf{Code} & \textbf{Description} \\
\hline
\code{SUCCESS} & \small{Indicates success of a lifecycle operation. Used in \code{PING}, \code{INIT}, \code{RECEIVE}, and \code{FETCH} operations.} \\
\code{FAILURE} & \small{Indicates that a requested operation failed. Used in \code{PING} and \code{INIT} operations.} \\
\code{MESSAGE} & \small{Signals that message carries a payload or key. Used in \code{SEND}, \code{POST}, \code{FETCH}, \code{RECEIVE}, and \code{FORWARD} operations.} \\
\code{NO\_MESSAGE} & \small{Signals that no payload was found for the requested node. Used in \code{FETCH} and \code{RECEIVE} operations.} \\
\hline
\end{tabular}
\end{table}

\paragraph*{Metadata}
For messages that are carrying a payload (those with \code{SEND}, \code{FETCH}, \code{POST}, \code{RECEIVE}, and \code{FORWARD} operations), the \code{Metadata} contains 3 sub-fields that hold information about the payload itself. The \textit{Source ID} (\code{src\_id}) and \textit{Destination ID} (\code{dest\_id}) sub-fields are free-form string identifiers that specify the source and destination of the payload, respectively.
Note that no naming convention is imposed and no translation is performed between application-layer and simulator-layer identifiers; therefore, the user must handle these identifiers appropriately at the endpoints.
The \textit{Payload Size} sub-field (\code{payload\_size}) specifies the size in bytes of the payload.
Even when a payload is cached in exchange for a payload key (see \S\ref{sec:framework-database}), this payload size should be retained as it  still refers to the original payload.

\paragraph*{Message Body}
The message body is a variable-length, versatile field that takes on different forms depending on the operation or payload status.
In payload-carrying messages, this field is the \textit{Payload} (\code{payload}) itself.
In those same messages, if the payload was cached (\S\ref{sec:framework-database}), then this field is the \textit{Payload Key} (\code{msg\_key}) for that payload.
During initialization (see \S\ref{sec:framework-system-init}), the body of an \code{INIT} message from a client to the daemon contains \textit{Introduction Details} (\code{IntroDetails}), which includes the client's identification and connection information.
The returned \code{INIT} message from the daemon to the client contains \textit{Configuration Parameters} (\code{ConfigParams}), which passes the system configuration to the newly connected client.

\subsection{Composition \& Extensibility}

NSB2 was designed as an open-source system so one of its primary design goals is to assure extensibility such that contributors can build improvements and usage-specific features atop the framework.
This manifests in our implementation as:
\begin{itemize}
    \item \textit{External extensibility.} The default implementation uses a curated technology stack of widely supported tools that themselves are extensible and modifiable.
    \item \textit{Internal modularity.} The implementation provides internal abstracted modular interfaces such that replacing a default stack component stays local to that component.
\end{itemize}

\subsubsection{Default Technology Stack}

With extensibility in mind, we present the components of the default technology stack and how it plays its role while advancing the design goals of being lightweight, accessible, and/or extensible:
\begin{itemize}
    \item \textbf{Transport} -- \textit{TCP Sockets (POSIX)}
    \newline Each sub-channel of the multi-channel client-to-daemon connection is realized as a TCP socket. Sockets introduce no broker process or external dependency, keeping the latency floor minimal and communication support uniform across platforms.

    \item \textbf{Wire Protocol} -- \textit{Protocol Buffers}
    \newline A single \code{.proto} schema is compiled to both C++ and Python bindings, providing a compact binary format and language-agnostic message definitions.
    This schema can be extended to incorporate additional features and fields, and new schemas can be implemented to accommodate new protocols.
    
    \item \textbf{Payload Cache} -- \textit{Redis}
    \newline The in-memory key-value store described in \S\ref{sec:framework-database} is implemented with \textit{Redis}.
    \textit{Redis} is fast, simple, and lightweight, and provides broad language support; it also provides options for the database to be accessed remotely should the NSB2 pipeline be distributed across machines.
    
    \item \textbf{Configuration Format} -- \textit{YAML}
    \newline The user-facing configuration file (\S\ref{sec:configuration}) is written in YAML, chosen because it is human-editable, dependency-light, and supported equivalently in both language bindings.
    
    \item \textbf{Build System \& Packaging} -- \textit{CMake} and \textit{pkg-config}
    \newline A single \textit{CMake} description targets MacOS, Linux, and Windows (via WSL); installed artifacts are registered with \textit{pkg-config} as \code{nsb} so dependent projects can locate NSB2 without bespoke build glue.
    
    \item \textbf{Implementation \& API Languages} -- \textit{C++17} and \textit{Python 3}
    \newline The daemon is implemented in C++17 while the endpoint client APIs are provided with bindings in both C++17 and Python 3, both widely used in the research and developer community.
    Because the client libraries are thin wrappers of system function calls to the daemon, API support can be extended to new languages with relative ease.
\end{itemize}

\subsubsection{Modular Interfaces} \label{sec:modular-interfaces}

The high-level implementations of the framework components do not reference any specific transport or cache implementation directly and instead use member modules built from abstract base classes, interacting with them via methods contracted in the base class and implemented in descendant module implementations.
Replacing these stack components can be done by simply writing a new module that derives from the appropriate base class and without changing the logic in the rest of the system.
We present two of these interfaces  -- \code{Comms} for transport and \code{DBConnector} for caching -- and how their operation signatures are realized in implementation.


\paragraph*{Transport Abstraction}
Following the communication architecture detailed in \S\ref{sec:framework-comms}, the \code{Comms} base class defines the shared \code{Channel} enumeration (\comms{CTRL}, \comms{SEND}, and \comms{RECV}) and a thin set of operations necessary for transport:
\begin{itemize}
    \item \code{connectToServer()} \newline \textit{\small to connect to a host}
    \item \code{sendMessage(channel, message)} \newline \textit{\small to transmit messages over a specified channel}
    \item \code{receiveMessage(channel, timeout)} \newline \textit{\small to receive messages over a specified channel}
    \item \code{listenForMessage(channel, timeout)} \newline \textit{\small to listen for messages asynchronously over a specified channel}
    \item \code{closeConnection()} \newline \textit{\small to close out the connection to the host}
\end{itemize}
While \code{SocketInterface} realizes these operations through parallel TCP connections to provide fast and minimal transport, a proof-of-concept \textit{RabbitMQ} interface has also been implemented, realizing these same operations using AMQP queues.
The implementation choice of transport is only local to the communication module and the rest of the code can remain unaffected.
Similarly, users can implement their own transport alternatives, such as shared memory for local experiments or RDMA for HPC environments.

\paragraph*{Payload Cache Abstraction}
Following the design detailed in \S\ref{sec:framework-database}, the \code{DBConnector} base class declares the contracted operations for caching payloads:
\begin{itemize}
    \item \code{store(value)}, returns \code{key} \newline \textit{\small write operation}
    \item \code{checkOut(key)}, returns \code{value} \newline \textit{\small destructive read operation}
    \item \code{peek(key)}, returns \code{value} \newline \textit{\small non-destructive read operation}
\end{itemize}
A shared key-generation helper is also provided in the base class to guarantee unique key generation.
While \code{RedisConnector} realizes these operations with \textit{Redis}, the contract is small enough that alternative key-value stores can be implemented, such as \textit{Memcached} for memory-constrained distributed deployments or file-backed storage for very large payloads.
Note that the use of any kind of external storage is optional and can be disabled in the configuration.

\paragraph*{Extension}
The NSB2 framework implementation \textit{requires} channel-segmented transport and key-value store and defines these needs through these shared contracts.
However, \textit{how} these requirements are met can be up to the users/contributors should they adapt NSB2 to their own unique preferences and constraints.
While the default technology stack aims to provide a general solution to minimize operational friction, modular interfaces provide flexibility and ease-of-integration when necessary, as validated in the implementation of the \textit{RabbitMQ} communication backend as an alternative to TCP sockets.

\subsection{Feature Configuration} \label{sec:configuration}

The user can configure NSB2 to best suit their application and system environments and needs.

\paragraph*{Daemon Connection Details}
The following parameters should be set to connect to the daemon:
\begin{itemize}
    \item \textbf{Daemon Address} -- \code{system.daemon\_address} \newline \textit{\small Address of the daemon.}
    \item \textbf{Daemon Port} -- \code{system.daemon\_port} \newline \textit{\small Port of the daemon.}
\end{itemize}

\paragraph*{\textbf{System Mode}}
The system mode can be configured with the following options:
\begin{itemize}
    \item \config{\textbf{PULL}} (\code{system.mode $\leftarrow$ 0})
    \newline {\small When set to \config{PULL} mode, the system will be polling-based, i.e., endpoint clients will have to poll the daemon to fetch and receive payloads. \config{PULL} mode is suitable for tightly-controlled environments and is a safe and portable option.}
    \item \config{\textbf{PUSH}} (\code{system.mode $\leftarrow$ 1})
    \newline {\small When set to \config{PUSH} mode, the system will automatically forward payloads from endpoint to endpoint, such that clients can listen for incoming payloads without explicitly polling for them. \config{PUSH} mode is recommended when lower latency is desired, when the environment is asynchronous, or endpoints are event-driven.}
\end{itemize}

\paragraph*{\textbf{Simulator Mode}}
The simulator mode can be configured with the following options:
\begin{itemize}
    \item \config{\textbf{system-wide}} (\code{system.simulator\_mode $\leftarrow$ 0})
    \newline {\small When set to \config{system-wide}, only one simulator client can exist in the NSB2 system and is responsible for managing all interactions with the network simulator. The \config{system-wide} simulator mode is aligned with top-down, centrally-modeled simulator options (e.g., \textit{ns-3}).}
    \item \config{\textbf{per-node}} (\code{system.simulator\_mode $\leftarrow$ 1})
    \newline {\small When set to \config{per-node}, the system assumes that one simulator client exists for each node in the network, with a corresponding node in the application space. The \config{per-node} simulator mode is aligned with bottom-up, module-level simulators (e.g., \textit{OMNeT++}).}
\end{itemize}

\paragraph*{Database Utilization Details}
To use the in-memory storage for more efficient payload relaying, the following parameters must be specified:
\begin{itemize}
    \item \textbf{Database Use} -- \code{database.use\_db} \newline \textit{\small Whether to use the database or not, either \code{true} or \code{false}.}
    \item \textbf{Database Address} -- \code{database.db\_address} \newline \textit{\small Address of the database.}
    \item \textbf{Database Port} -- \code{database.db\_port} \newline \textit{\small Port of the database.}
    \item \textbf{Database Number} -- \code{database.db\_num} \newline \textit{\small The specific index of the database instance.}
\end{itemize}

The system configuration defines how the infrastructure is constructed and operated and how the components behave, as we discuss throughout the rest of this paper.

\subsection{Framework Operation}

The operational life of the NSB2 framework involves initializing the system, processing payloads for the duration of the co-simulation, and halting the system.
Throughout operation, the daemon is the central, long-running service that keeps the system functioning; other components of the stack, like the transport and the payload cache, must also remain active and reachable.
The endpoint clients, however, are ephemeral and can join and leave the framework as necessary without critically affecting the system as a whole.
Figures~\ref{fig:init-handshake}, \ref{fig:timing-pull}, and \ref{fig:timing-push} provide timing diagrams of the processes described in this section.

\subsubsection{Initialization} \label{sec:framework-system-init}

Prior to initializing the system, the external cache and any transport dependencies must be ready for use.
At initialization, the daemon starts up and parses the user-defined configuration file to gather specifications (preserving them to disseminate to the clients) and sets its internal state (e.g., the client registry and TX/RX buffers).
It binds to the specified port and starts listening for incoming connections, ready to service \code{INIT} requests from clients.

\begin{figure}
    \centering
    \includegraphics[width=\linewidth]{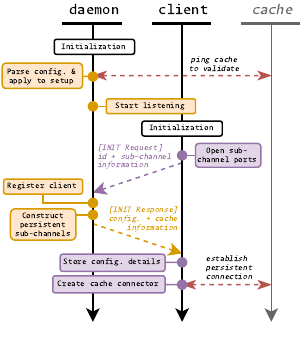}
    \caption{Initialization handshake.}
    \label{fig:init-handshake}
\end{figure}

\paragraph*{Client Initialization}
Each logical node in the application space must be initialized with its own \code{NsbAppClient} instance.
If the simulator mode is configured to be \config{system-wide}, then only a single \code{NsbSimClient} instance should be instantiated at the control plane of the network simulator, e.g., the main simulator coordinstor or event scheduler.
If it is configured to be \config{per-node}, then each logical node in the simulated network must be initialized with its own \code{NsbSimClient} instance.
Upon construction, each client initializes into the framework with a handshake with the daemon as follows:
\begin{enumerate}
    \item The client constructs its communication sub-channels, then uses a temporary connection to contact the daemon at the specified address and port and transmit an \code{INIT} message with its identifier and sub-channel port information.
    \item The daemon parses this \code{INIT} message, saves the client information, and constructs persistent sub-channel connections to be used from then onward to communicate with that client. The daemon then responds to the client with its own \code{INIT} message with the configuration parameters, ensuring that the same configuration is understood by the daemon and the endpoints.
    \item The client receives the daemon's response and saves the enclosed configuration parameters. The client then connects to the payload cache with the address and port specified in the configuration.
\end{enumerate}
Figure~\ref{fig:init-handshake} shows a visual representation of this process.
After initialization, the clients are fully registered with the daemon, communications are live, and the system is ready to start processing payloads.

\subsubsection{Lifecycle of a Payload} \label{sec:framework-lifecycle}

\begin{figure}
    \centering
    \includegraphics[width=\linewidth]{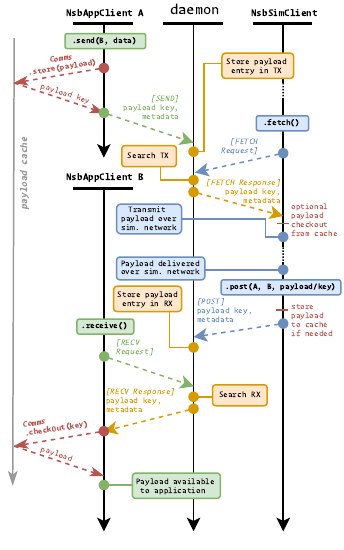}
    \caption{Payload processing with system in \config{PULL} mode.}
    \label{fig:timing-pull}
\end{figure}
\begin{figure}
    \centering
    \includegraphics[width=\linewidth]{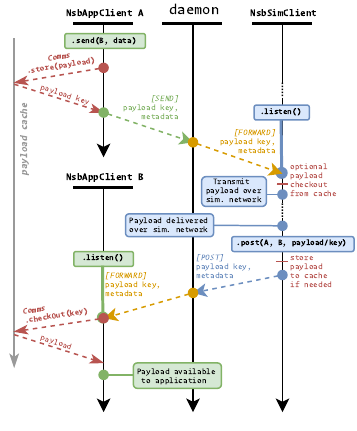}
    \caption{Payload processing with system in \config{PUSH} mode.}
    \label{fig:timing-push}
\end{figure}

The initialized framework is ready to facilitate the pipeline in which payloads are relayed from the sending application through a simulated network to the receiving application.
This is achieved through 4 main steps, with each step being of point of API integration:
\begin{enumerate}
    \item \textbf{Sending the payload from the sender node.} \newline
    \textit{\small \code{NsbAppClient.send} implemented in TX logic of application.} \newline
    The sending application client caches the payload in the database and encloses the payload key and metadata in a \code{SEND} message to transmit to the daemon.
    Upon reception, the daemon creates and stores the payload entry in its transmission (TX) buffer.
    \item \textbf{Fetching \& relaying the payload through the network.} \newline
    \textit{\small \code{NsbSimClient.fetch} implemented in TX logic of simulator.} \newline
    The simulator client awaits or polls for new payloads.
    If the system is in \config{PULL} mode, then the simulator client sends a \code{FETCH} message, optionally specifying which node it is fetching for.
    The daemon searches its TX buffer and responds with either the requested payload entry or a \code{NO\_MESSAGE} message.
    If the system is in \config{PUSH} mode, the payload entry will be forwarded immediately from the preceding \code{SEND} message.
    Once a payload entry is retrieved, the simulator client uses the payload key to check out the full payload and passes it onto the network simulator to be transmitted through the simulated network.
    The user may optionally choose not to check out the full payload and instead construct a proxy payload using the payload metadata.
    \item \textbf{Posting the payload if it is successfully delivered.} \newline
    \textit{\small \code{NsbSimClient.post} implemented in delivery logic of simulator.} \newline
    Upon successful delivery of the payload in the simulated network, the simulator client caches the payload to the database sends a \code{POST} message containing the payload key and payload metadata to the daemon.
    Upon reception, the daemon creates and stores the payload entry in its reception (RX) buffer.
    \item \textbf{Receiving the payload from the receiver node.} \newline
    \textit{\small \code{NsbAppClient.receive} implemented in RX logic of application.} \newline
    The receiving application client awaits or polls of newly received payloads.
    If the system is in PULL mode, then the application client sends a \code{RECEIVE} message to the daemon, optionally specifying which node it is requesting on behalf of.
    The daemon searches its RX buffer and responds with either the requested payload entry or a \code{NO\_MESSAGE} message. If the system is in PUSH mode, the payload entry will be forwarded immediately from the preceding \code{POST} message. Once a payload entry is retrieved, the application client uses the payload key to check out the full payload and passes it onto the application.
\end{enumerate}
This process is illustrated in Figure~\ref{fig:lifecycle} with the main steps being shown as their respective API calls and Figure~\ref{fig:timing-diagrams} shows the timing in \config{PULL} (Fig.~\ref{fig:timing-pull}) and \config{PUSH} (Fig.~\ref{fig:timing-push}) system configurations.
Communication details can be found in Table~\ref{tab:protobuf-manifest-op}.
Snippet~\ref{lst:int-app-python} shows an example application-side NSB2 integration through \code{NsbAppClient}.
Snippet~\ref{lst:int-sim-cpp} shows an example simulator-side NSB2 integration through \code{NsbSimClient}, very similar to how it would be implemented in \textit{OMNeT++}.

\begin{lstlisting}[float, style=nsb, language=Python, label={lst:int-app-python}, caption={Application-side integration in Python.}]
import nsb_client as nsb
class ApplicationNode:
    def __init__(self, ...):
        self.conn = nsb.NSBAppClient(self.name,
                                     "127.0.0.1", 65432)
    def send_data(self, destination):
        data = ...  # Prepare data as bytes.
        self.conn.send(destination, data)
    def receive_data(self):
        msg = self.conn.receive()
        if msg:
            data = msg.payload
            # Process message.
            ...\end{lstlisting}

\begin{lstlisting}[float, style=nsb, language=C++, label={lst:int-sim-cpp}, caption={Simulator-side integration in C++.}]
#include "nsb_client.h"
class SimulatorNode {
private:
    std::string name;
    nsb::NSBSimClient conn;
public:
    SimulatorNode(...) {
        conn = nsb::NSBSimClient(name, "127.0.0.1", 65432);
        ...
    void sendMessage(...) {
        nsb::MessageEntry msg = conn.fetch();
        if (msg.exists()) {
            std::string payload = msg.payload_obj;
            // Parse and transmit message.
            ...
    void handleMessage(...) {
        // If payload has reached destination...
        if (...) {
            conn.post(source, destination, payload);
            ...\end{lstlisting}

\subsubsection{Shutdown}
Any one of the framework components -- the daemon or the clients -- can send a shutdown signal by transmitting an \code{EXIT} message to the daemon, which will then disseminate this message throughout the system.
The daemon and clients will then proceed to tear down the transport connections and the clients will terminate their connection to the cache.
\section{Performance Evaluation} \label{sec:evaluation}
To evaluate NSB2's overhead and scalability, we conducted a series of controlled stress tests with the system configuration in PULL mode to showcase the upper bounds of latency and resource utilization.
Each experiment was run for 60 seconds and repeated across five independent sweeps, averaged in the results shown below.
The averaged aggregate dataset across those five sweeps is presented in this section.

\subsection{Experimental Setup}

Our evaluation matrix consisted of practical node pool sizes of 10, 50, and 100 nodes and an extreme case of 500 nodes; each node size configurations produced  aggregate offered loads of 10, 100, and 1000 messages per second (msg/s).
The configured message rate is the total offered load system-wide across the experiment.
The experiments used a lightweight ghost simulator client that immediately returned fetched payloads to the framework, allowing us to isolate bridge-side transport, queueing, and relay behavior without conflating the measurements with the execution cost of a detailed network model or network simulator processing.

\paragraph{Machine Specifications}
We evaluated the system in an isolated computing environment provided by Cloudlab~\cite{cloudlab}.
The machine specifications are shown below in Table~\ref{tab:machine}.
\begin{table}[!ht]
    \centering
    \begin{tabular}{|c|l|}
        \hline
        \textbf{\small CPU} & AMD EPYC 7402P \textit{\small (24 Cores, 48 Threads)} \\ \hline
        \textbf{\small Architecture} & x86\_64 \textit{\small (Little Endian)} \\ \hline
        \textbf{\small Virtualization} & AMD-V Supported \\ \hline
        \textbf{\small RAM} & 125 GB Total \textit{\small (122 GB Available)} \\ \hline
        \textbf{\small Swap} & 8.0 GB Total \\ \hline
    \end{tabular}
    \caption{Test machine specifications.}
    \label{tab:machine}
\end{table}

\paragraph*{Metrics}
To profile latency across the co-simulation path, we instrumented the framework by extending the protocol buffer message definitions (see \S\ref{sec:framework-messaging}) to capture several timing metrics as each payload travels across the bridge.
We used this information to observe latency through the time spent resident inside NSB only, computed as the outbound daemon-resident interval plus the return daemon-resident interval, and excluding the interval during which the simulator client holds the payload outside the daemon.
In addition to latency, we observed daemon CPU utilization and peak memory usage through the resident set size (RSS) during each run to capture resource overhead.

\subsection{Results \& Discussion}

\begin{figure*}[t]
%
%
%

\pgfplotstableread[col sep=comma]{plots/eval_data.csv}\nsbevaldata

\pgfplotsset{
    nsb cluster/.style={
        ybar=2pt,
        bar width=6pt,
        width=\linewidth,
        height=5.2cm,
        symbolic x coords={10,50,100,500},
        xtick=data,
        enlarge x limits=0.12,
        nodes near coords,
        nodes near coords style={font=\tiny, anchor=west},
        every node near coord/.append style={yshift=1pt},
        every axis plot/.append style={fill opacity=0.92, draw opacity=0},
        tick style={draw=none},
        axis lines*=left,
        xlabel={\# Nodes},
    },
}


{\centering
\begin{tikzpicture}[
    swatch/.style={draw=none, minimum width=11pt, minimum height=6pt, inner sep=0pt},
    lab/.style={font=\footnotesize, anchor=west, inner xsep=4pt},
]
\node[swatch, fill=blue!55!cyan]                (s1) at (0,0)                  {};
\node[lab]                                      (l1) at (s1.east)              {10 msg/s};
\node[swatch, fill=orange!85!red, anchor=west]  (s2) at ([xshift=12pt]l1.east) {};
\node[lab]                                      (l2) at (s2.east)              {100 msg/s};
\node[swatch, fill=teal!75!black, anchor=west]  (s3) at ([xshift=12pt]l2.east) {};
\node[lab]                                            at (s3.east)             {1000 msg/s};
\end{tikzpicture}\par}
\vspace{2pt}

\begin{subfigure}[t]{0.33\textwidth}
\centering
\begin{tikzpicture}
\begin{axis}[
    nsb cluster,
    ymin=0, ymax=260,
    ytick={0,50,100,150,200,250},
    ylabel={Avg.\ latency (ms)},
    ymajorgrids, grid style={dotted, gray!40},
    nodes near coords style={font=\tiny, rotate=90, anchor=west,
        /pgf/number format/.cd, fixed, precision=1, /tikz/.cd},
]
\addplot[fill=blue!55!cyan]  table [x=nodes,
    y expr=\thisrow{nsb_residence_avg_s}*1000,
    restrict expr to domain={\thisrow{rate}}{10:10}]   {\nsbevaldata};
\addplot[fill=orange!85!red] table [x=nodes,
    y expr=\thisrow{nsb_residence_avg_s}*1000,
    restrict expr to domain={\thisrow{rate}}{100:100}] {\nsbevaldata};
\addplot[fill=teal!75!black] table [x=nodes,
    y expr=\thisrow{nsb_residence_avg_s}*1000,
    restrict expr to domain={\thisrow{rate}}{1000:1000}]{\nsbevaldata};
\end{axis}
\end{tikzpicture}
\caption{NSB2-incurred round-trip latency.}
\label{fig:eval-latency}
\end{subfigure}\hfill
\begin{subfigure}[t]{0.33\textwidth}
\centering
\begin{tikzpicture}
\begin{axis}[
    nsb cluster,
    ymin=0, ymax=100,
    ytick={0,20,40,60,80,100},
    ylabel={Avg.\ daemon CPU (\%)},
    ymajorgrids, grid style={dotted, gray!40},
    nodes near coords style={font=\tiny, rotate=90, anchor=west,
        /pgf/number format/.cd, fixed, precision=1, /tikz/.cd},
]
\addplot[fill=blue!55!cyan]  table [x=nodes, y=avg_cpu_percent,
    restrict expr to domain={\thisrow{rate}}{10:10}]   {\nsbevaldata};
\addplot[fill=orange!85!red] table [x=nodes, y=avg_cpu_percent,
    restrict expr to domain={\thisrow{rate}}{100:100}] {\nsbevaldata};
\addplot[fill=teal!75!black] table [x=nodes, y=avg_cpu_percent,
    restrict expr to domain={\thisrow{rate}}{1000:1000}]{\nsbevaldata};
\end{axis}
\end{tikzpicture}
\caption{Daemon CPU utilization.}
\label{fig:eval-cpu}
\end{subfigure}\hfill
\begin{subfigure}[t]{0.33\textwidth}
\centering
\begin{tikzpicture}
\begin{axis}[
    nsb cluster,
    ymin=0.0, ymax=8.0,
    ytick={2.0,4.0,6.0,8.0},
    ylabel={Memory usage (MB)},
    ymajorgrids, grid style={dotted, gray!40},
    nodes near coords style={font=\tiny, rotate=90, anchor=west,
        /pgf/number format/.cd, fixed, precision=2, /tikz/.cd},
]
\addplot[fill=blue!55!cyan]  table [x=nodes, y=peak_memory_mb,
    restrict expr to domain={\thisrow{rate}}{10:10}]   {\nsbevaldata};
\addplot[fill=orange!85!red] table [x=nodes, y=peak_memory_mb,
    restrict expr to domain={\thisrow{rate}}{100:100}] {\nsbevaldata};
\addplot[fill=teal!75!black] table [x=nodes, y=peak_memory_mb,
    restrict expr to domain={\thisrow{rate}}{1000:1000}]{\nsbevaldata};
\end{axis}
\end{tikzpicture}
\caption{Peak daemon memory usage.}
\label{fig:eval-memory}
\end{subfigure}
    \caption{NSB2 performance evaluation under increasing node count and system-wide message rate.}
    \label{fig:eval}
\end{figure*}
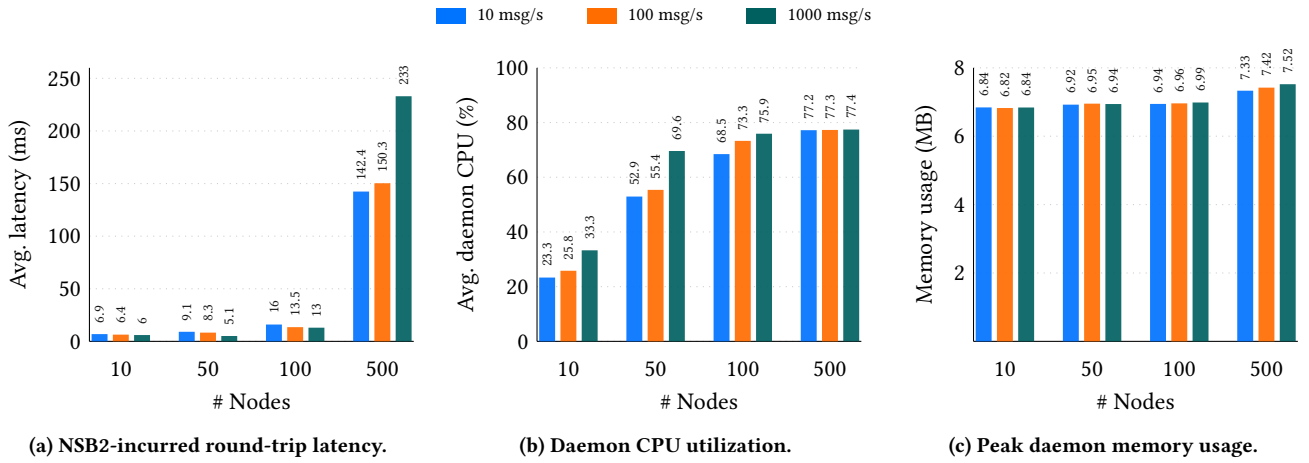

Our results are shown in Figure~\ref{fig:eval}, with latency shown in Figure~\ref{fig:eval-latency}, daemon CPU utilization shown in Figure~\ref{fig:eval-cpu}, and peak daemon memory usage shown in Figure~\ref{fig:eval-memory}.

\paragraph*{Practical Usage}
NSB2 introduces relatively low latency (shown in Figure~\ref{fig:eval-latency}) across practical co-simulation regimes ($\leq$100 nodes at all payload transmission rates), with CPU utilization below 76\%, and memory usage under 7 MB.
As expected, CPU utilization scales up as the number of nodes increases, latency increases only sub-linearly, and memory usage stays relatively consistent.
We observe that the average latency actually goes down as the transmission rate increases; this is consistent with connection time and system warm-up amortizing across more payloads.
Payload transmission rate does have a minor impact on CPU utilization, which scales sub-linearly with the rate.
Overall, performance for 10--100 nodes remains lightweight and stable for all payload transmission rates.

\paragraph*{Upper Bounds on Usage}
The main departure from this behavior occurs at 500 nodes, average latency skyrockets to 142.4 ms, 150.3 ms, and 233.0 ms for 10 msg/s, 100 msg/s, and 1000 msg/s transmission rates and average daemon CPU utilization caps out at \~77\%.
These observations indicate a limiting factor of our system and operating environment -- the daemon leverages the environment's file descriptors to create NSB2's persistent communication architecture (see \S\ref{sec:framework-comms}), creating 3 sub-channels per each of 500 nodes, resulting in the monitoring of 1500 file descriptors.
This exceeds the file descriptor set size of our operating environment which is 1024 such that the current polling implementation within the daemon must negotiate and reconstruct the file descriptor set, adding latency in each individual payload path and increasing CPU utilization in the daemon.
Note that it is this single-threaded polling mechanism that contributes to performance degradation rather than the per-payload processing cost.
Memory usage in these conditions only experiences a very slight increase from $<$7 MB to 7.5 MB.
These observations present an opportunity to improve the daemon to manage such environmental constraints more efficiently and thus make NSB2 more robust and capable of handling larger networks even in constrained environments, e.g., new polling mechanisms within the daemon to more efficiently handle the file descriptor set size limitation.

\section{Conclusion} \label{sec:conclusion}

In this paper, we presented NSB2, an open-source application-network co-simulation middleware that consists of a central daemon and thin application- and simulator-side libraries that realy payloads between users' application processes and a network simulator backend.
NSB2 was designed and engineered to be lightweight, accessible, and extensible, and realized these goals through an extensible technology stack, sub-channeled communication architecture, and internal modular abstractions.
Our performance evaluation shows that in practical usage scenarios, NSB2 is indeed lightweight and fast, with minimal bridge latency and sub-7 MB peak memory usage.
Overall, NSB2 aims to lower the cost of network co-simulation in complex systems modeling, making the network component a first-class part of connected applications and systems development and making joint application-network research and development easier and more accessible.

\paragraph*{Ongoing Work \& Future Directions}
As our evaluation also exposes an environment constraint affecting the daemon's polling and message handling that poses a ceiling on bridge performance in more extreme usage scenarios, we aim to make our communication architecture and daemon polling mechanisms more adaptable to such constraints in the face of more intense usage scenarios.
Beyond our current implementation, we also want to refine NSB2 for multi-machine deployments, a capability that should be possible with its current communication and data infrastructure but one that has not been fully tested and optimized for.
In the development of these improvements and new features, we make use of NSB2's modularity to update components of the overall stack individually, without changing the usage of the bridge endpoints.
We also invite contributors from networking-related domains to build their own domain-specific adaptations and features.

\paragraph*{Open-Source Repository}
NSB2 is open-source and available through our public repository\footnote{\textit{Public repository link omitted due to double-blind submission requirement.}}.
It contains the source code, installation tools, documentation, and integration templates for \textit{OMNeT++} and \textit{ns-3}.
The project is open and readily available to new users and contributors who want to build upon and improve NSB2 for all users or even just users within their own domain.

\bibliographystyle{ACM-Reference-Format}
\bibliography{references}

\end{document}